\documentclass[paper]{JHEP3}
\usepackage{graphicx}   
\usepackage{picins}   
\newcommand{\nnb}{\nonumber}
\newcommand{\be}{\begin{eqnarray}}
\newcommand{\ee}{\end{eqnarray}}
\newcommand{\bea}{\begin{eqnarray}}
\newcommand{\eea}{\end{eqnarray}}
\newcommand{\barl}{\begin{array}{rl}}
\newcommand{\ba}{\begin{array}}
\newcommand{\ea}{\end{array}}

\title{Matrix Elements and Parton Showers in Hadronic Interactions}
\author{F. Krauss\\
        Institut f{\"u}r Theoretische Physik,\\
        TU Dresden\\
        D-01062 Dresden, Germany\\
        E-mail: krauss@theory.phy.tu-dresden.de}
\abstract{
\noindent A method is suggested to combine tree level QCD matrix 
for the production of multi jet final states and the parton shower 
in hadronic interactions. The method follows closely an algorithm 
developed recently for the case of $e^+e^-$ annihilations 
\cite{Catani:2001cc}.}


\begin{document}

\section{Introduction: ME vs. PS}
The production of multi jet final states is one of the searching grounds 
for new physics at current and future collider experiments. Therefore
their Monte Carlo simulation, both for signal and background 
processes is of crucial importance for the success of the experiments.
Currently, two alternative ways to fill the phase space for particle
emission can be formulated:
\begin{enumerate}
\item One might use exact matrix elements (ME) at some given 
      perturbative order in the coupling constant(s), say $\alpha_s$.
      In moxst cases, these MEs are available at the tree level,
      i.e.\relax \  at the lowest perturbative order for the production of 
      a specific final state. Then, the final state particles are 
      usually identified with jets and their phase space is cut
      accordingly. In some cases higher perturbative orders are known 
      as well, and blur this simple correspondence. However, in this 
      paper we will focus on tree level MEs, therefore we can stick to
      the simple identification of partons with jets. The virtue 
      of the MEs is that they are exact and take {\bf all} interference 
      effects at the perturbative order into account. The downside is 
      that more and more diagrams have to be considered with rising 
      numbers of outgoing particles populating a multidimensional
      phase space. Clearly, our calculational abilities do not live
      up to this situation. Furthermore, for the transition of the 
      partons into hadrons a phenomenological fragmentation model has
      to be invoked with parameters, that are {\it a priori} free and
      depend on the fragmentation scale. Hence, applying a
      fragmentation scheme immediately at the hard scale of the ME  
      necessitates a re-tuning of its parameters for {\bf each}
      c.m. energy in order to gain reliable results on the hadron
      level. This is an impossible task and limits the applicability
      of using pure MEs considerably.  
\item Alternatively one might use the parton shower (PS). By
      expanding around the soft and collinear limits of parton 
      emission the radiation pattern factorises into individual parton
      branchings, each giving rise to potentially large soft and
      collinear logarithms. Via multiple emission according to the
      individual branching probabilities the PS then resums all the large 
      logarithms. In fact, the coherent, angular ordered PS is correct
      up to Next--to Leading Logarithmic (NLL) accuracy, i.e.\relax \ 
      it correctly resums {\bf all} terms of the form 
      $\alpha_s^n \left(L^{2n} +  L^{2n-1}\right)$, where $L$ denotes
      a large logarithm of the form $\ln s/Q_0^2$ with $s$ the hard
      scale of the process and $Q_0^2$ the soft parton resolution
      scale. Obviously, the PS is able to connect both the hard
      scale and the fragmentation scale. Hence, using the PS allows
      us to tune the fragmentation parameters at one scale, say at LEP1
      energies, where huge statistics were amassed, and to use the
      same parameters at {\bf all} other scales. However, due to its
      factorised structure the PS misses important interference
      effects that may play a significant role for detailled analyses. 
\end{enumerate} 
To gain some insight into the effect of these interference effects,
let us compare the emission of a gluon in $e^+e^-\to q\bar q$.
Pictorially we can write:
\begin{figure}[h]
\begin{tabular}{cc}
\begin{minipage}[ht]{8cm} {
\bea
\frac{d\sigma_{\rm ME}}{dx_1dx_2} &\sim& 
\left| 
\unitlength 1mm
\begin{picture}(28,12)
\put(0,-8){\includegraphics[width=1cm,height=2cm]{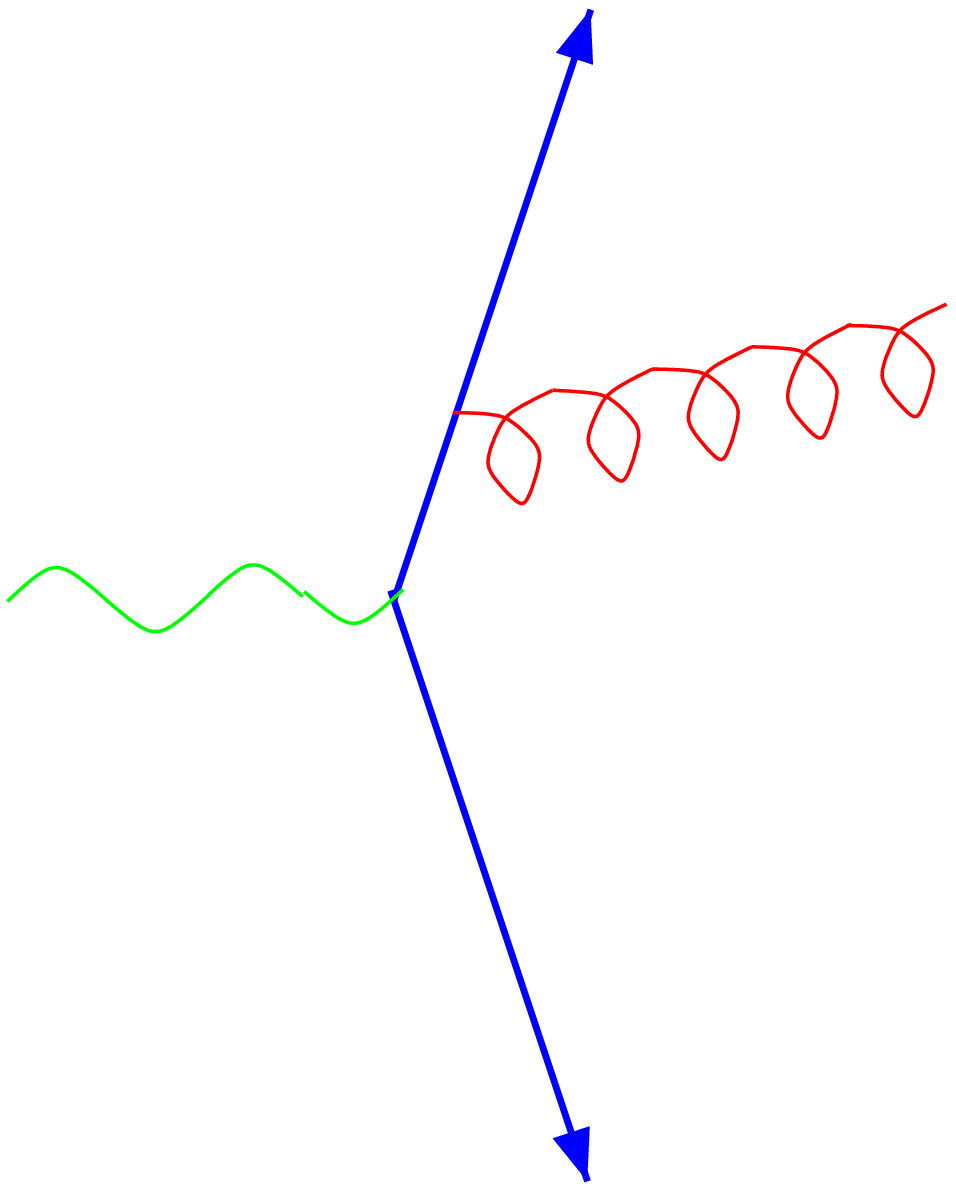}}
\put(11.5,2.0){+}
\put(15,-8){\includegraphics[width=1cm,height=2cm]{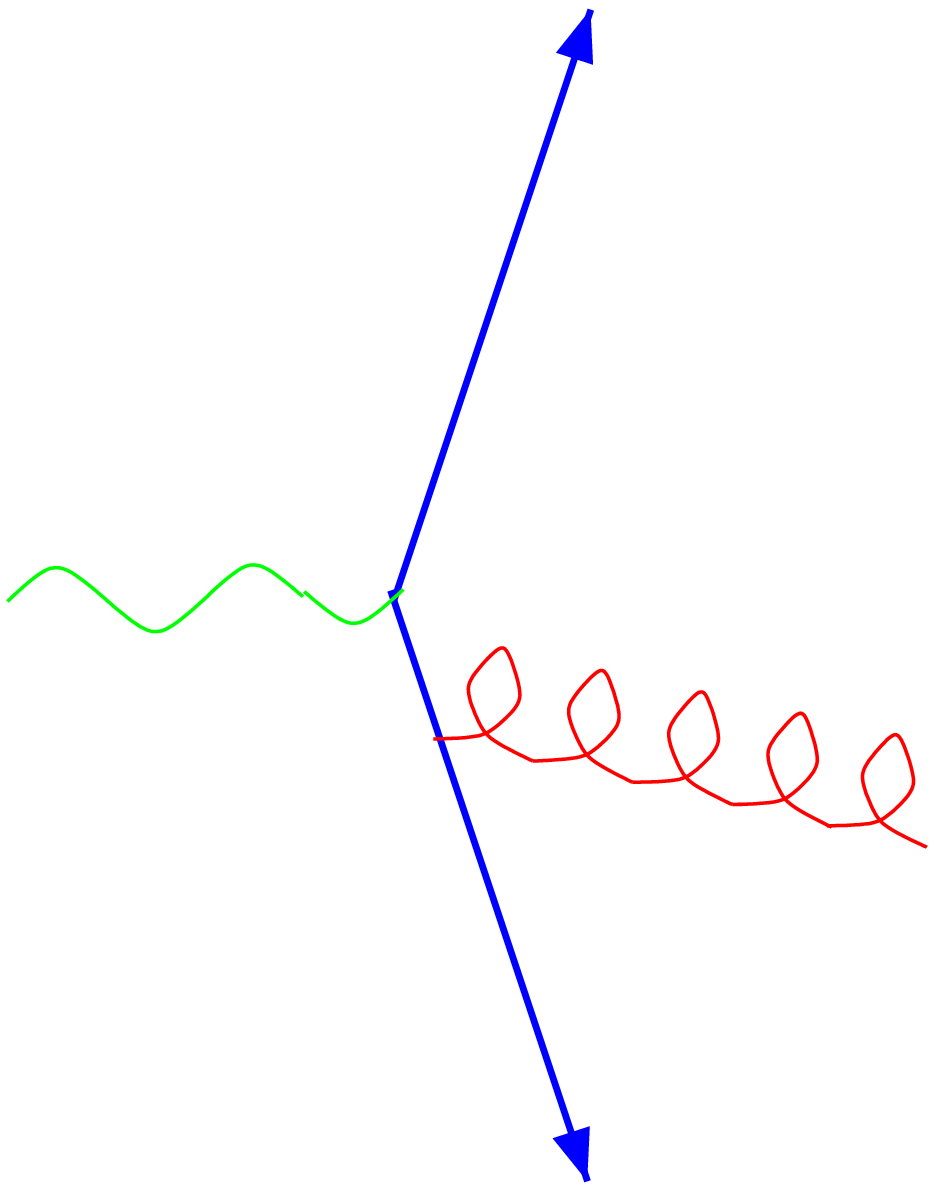}}
\end{picture}
\right|^2\nnb\\
\frac{d\sigma_{\rm PS}}{dx_1dx_2} &\sim&
\unitlength 1mm
\left|
\begin{picture}(10,12)
\put(0,-8){\includegraphics[width=1cm,height=2cm]{3j1.eps}}
\end{picture}\;
\right|^2
+
\left|
\begin{picture}(10,12)
\put(0,-8){\includegraphics[width=1cm,height=2cm]{3j2.eps}}
\end{picture}\;
\right|^2\nnb
\eea
}\end{minipage}
&
\begin{minipage}[ht]{5cm} {
\includegraphics[width=5cm,height=5cm]{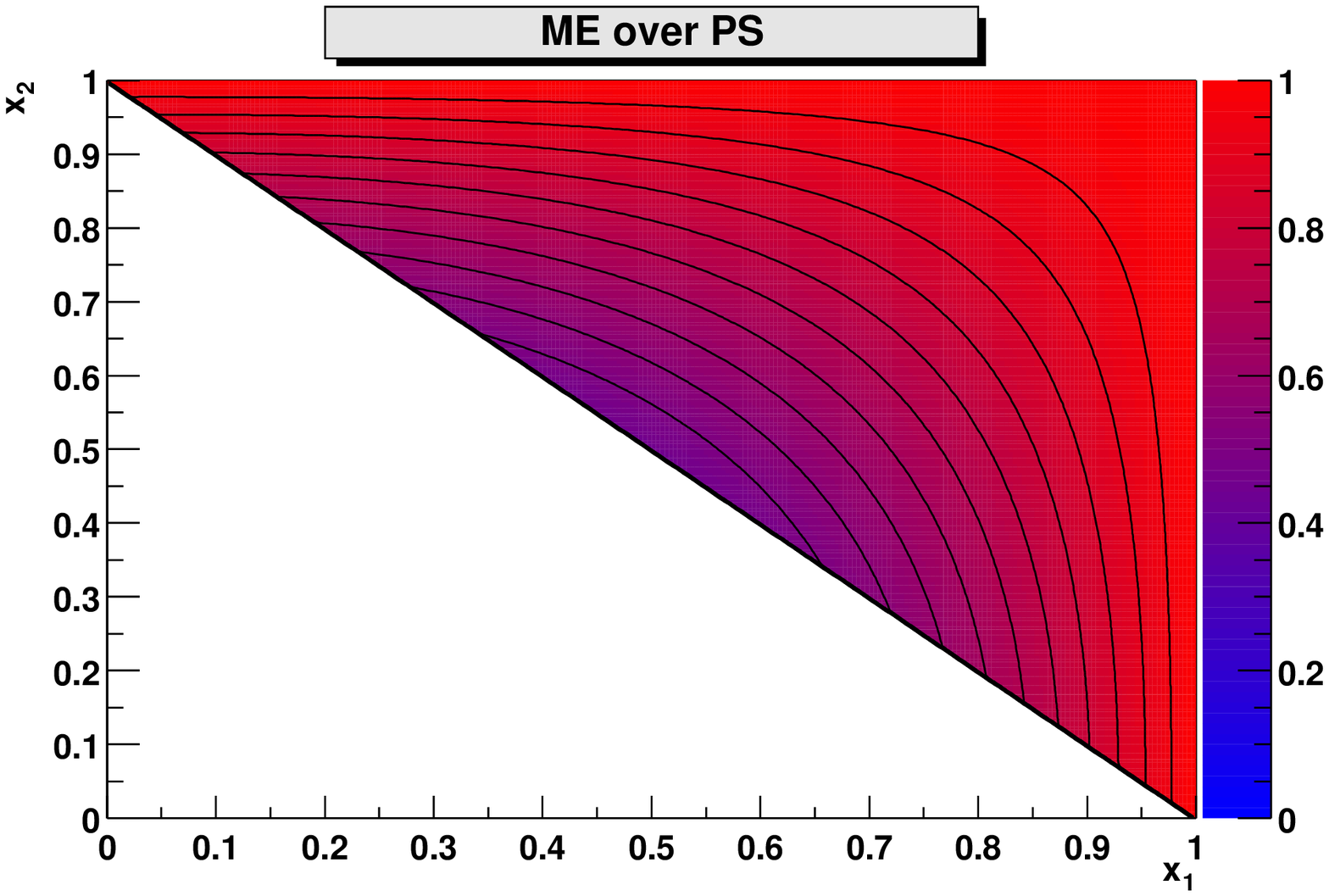}
}\end{minipage}
\end{tabular}
\caption{
Pictorial representation of contributions to single gluon emission in
$e^+e^-\to q\bar qg$ as given by the ME and the PS. As indicated, the
PS does not take into account the interference contributions present
in the ME. This leads to the ratio $ME/PS$ as depicted in the contour
plot. At the soft and collinear boundaries of the phase space in the
$x_1$-$x_2$ plane (where $x_i$ corresponds to the energy fraction of 
the quark and antiquark) the PS correctly reproduces the ME, whereas
in the region of hard gluon emission the PS omits the (destructive)
interference contribution.}
\end{figure}

From the figure above, it can be seen clearly, that in the ratio
$d\sigma_{\rm ME}/d\sigma_{\rm PS}$ the omitted interference terms
have a large impact in the region where the gluon is hard and emitted
at a large angle. Obviously it would be of great interest to
combine both ME and PS to take full advantage of their respective
virtues. A number of attempts in this direction were made in 
\cite{Belyaev:2000wn}-\cite{Dobbs:2001gb}, some of them also incorporate
NLO corrections for specific processes. For the latter, a recent paper
\cite{Frixione:2002ik} provides a general method to construct process
specific algorithms to combine MEs at NLO with the PS. This method,
however, is limited by the number of processes that are calculated at
NLO and that the user wants to implement. 

In contrast, in the framework of this paper I would like to restrict
the discussion on the merging of MEs at tree level with the PS. This
allows for a process independent method that can be applied
immediately. In fact, what I propose in this paper is the extension of
a previous paper \cite{Catani:2001cc}, addressing the merging of
arbitrary MEs for jet production at tree level and the PS in $e^+e^-$
annihilations, to processes with hadronic initial states. 

\parpic[r]{
\unitlength 1mm
\begin{picture}(70,120)
\put(0,35){\includegraphics[width=7cm]{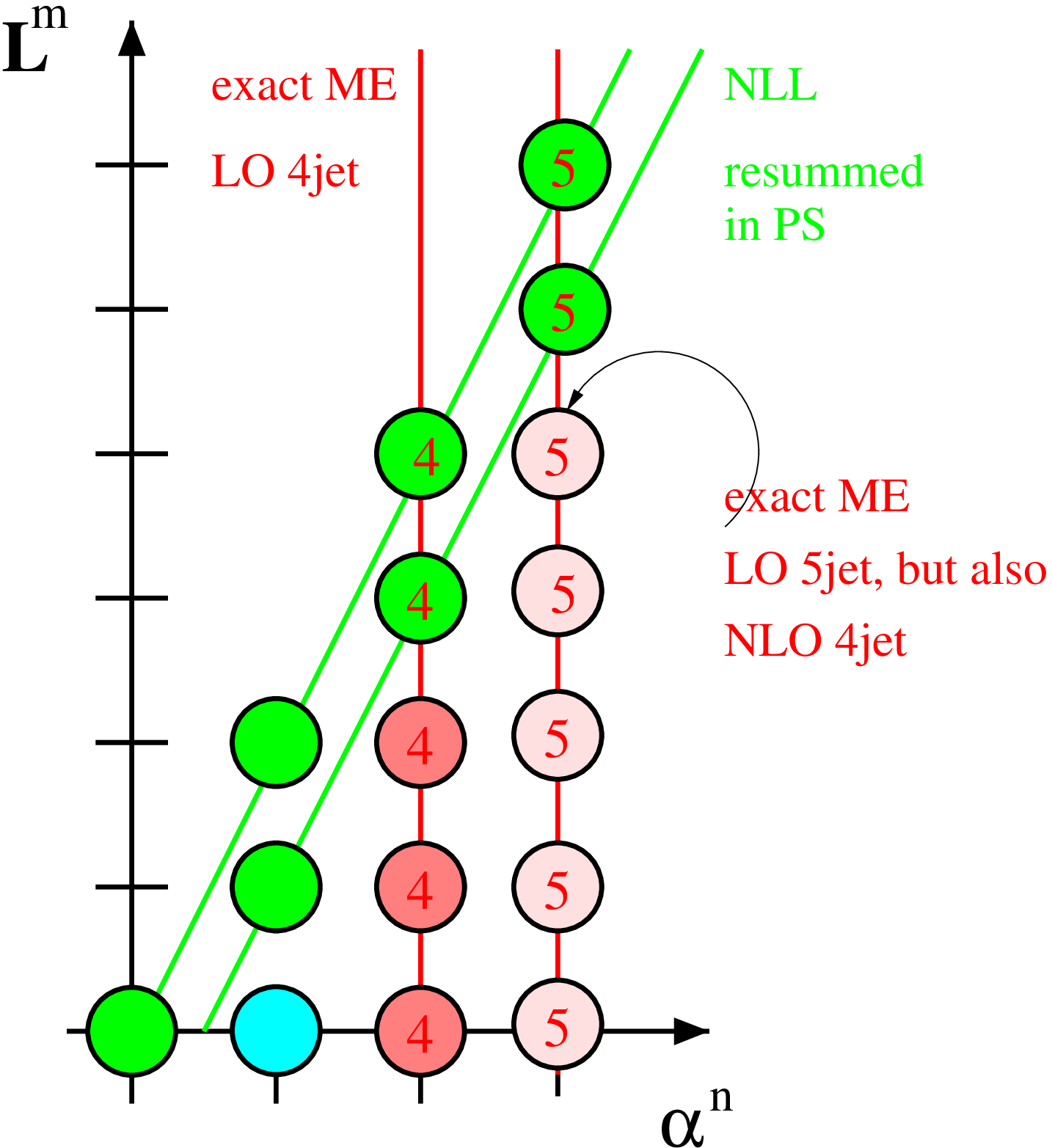}}
\put(0,15){
\begin{minipage}[ht]{7cm}
Schematic picture of perturbative orders in $e^+e^-\to $jets.
Clearly, for every order in $\alpha_s$ one additional gluon can be
emitted that can be soft or collinear leading to maximal two more
large logaritms.
\end{minipage}}
\end{picture}
} 
The basic idea of this method is to divide the phase space for
parton radiation in two disjoint regions, one region of jet production
that is filled with help of MEs and a complementary region of jet
production, that is filled with the PS. The combined ME+PS then should
correctly reproduce all large logarithms up to NLL accuracy {\bf and}
the sub-leading terms present in the ME. Consider as an example the 
process $e^+e^-\to$ jets. The respective orders in $\alpha_s$ and
$L$ are given in the figure to the right. For the case of four jet
production, the method incorporates all the contributions labelled
with a 4, i.e.\relax \  the full tree level ME, {\bf plus all} NLL
contributions of higher perturbative order, represented by the green 
blobs to the right of the vertical four jet line.

The challenge in such a procedure is to avoid double counting. In our
example of four jet production this translates into ensuring that the 
green blobs labelled by a 4 are taken into account only once.
This can be achieved by applying weights on the tree level ME such
that the NLL terms of higher perturbative order in $\alpha_s$ are
correctly treated and by a veto on the production of extra jets in the
PS. To see how this works, let us stick first to the example of jet
production in $e^+e^-$ annihilations before going to more general
cases.  

\section{ME+PS in $e^+e^-$ annihilations}

\subsection{Jet rates at NLL accuracy}
I'd like to begin the discussion of the combination procedure with a
brief review of the PS. There, individual branchings, i.e.\relax \  parton
emissions, are governed by the Sudakov form factor,
\bea\label{SudDef}
\Delta(T,t) = 
\exp\left\{-\int\limits_t^T\displaystyle{\frac{dt'}{t'}}
      \int\limits_{z_-(t')}^{z_+(t')} dz
      \displaystyle{\frac{\alpha[p_\perp(z,t')]}{2\pi}}\,
            P_{a\to bc}(z)\right\}\,,
\eea
yielding a {\bf probability for no branch} of parton $a$ into partons
$b$ and $c$ between scales $T$ and $t$. In Eq.(\ref{SudDef}) the
scales $T$ and $t$ might denote virtual masses like for instance in
PYTHIA \cite{Sjostrand:2000wi} or scaled opening angles like in
HERWIG, \cite{Marchesini:1992ch}. 
$z$ denotes the energy fraction of the parton after branching,
the limits on the $z$--integration depend on the meaning of the scale
parameters as does the particular form of the transverse momentum
$p_\perp$ in the argument of $\alpha_s$, and the splitting function
$P_{a\to bc}$ depends on the parton types involved. In the following
we will consider only angular ordered, coherent PS
\cite{Marchesini:1988cf}. Then the scales $T$ and $t$ play the role of
scaled opening angles, i.e.\relax \  the branching scale of parton $a$, $t_a$
reads 
\bea
t_a = E_a^2(1-\cos\theta_{bc})
\eea
and the $z$--limits are given by
\bea
\sqrt{t_0/t_a} < z < 1-\sqrt{t_0/t_a}
\eea
where $t_0$ is connected to a minimal branching angle. 

However, up to NLL accuracy we can then perform the $z$--integrals in
Eq.(\ref{SudDef}) yielding the NLL--Sudakov form factor
\bea\label{SudNLLDef}
\Delta_{q,g}^{\rm NLL}(Q,q) = 
\exp\left[-\int\limits_q^Q dq' \Gamma_{q,g}(q',Q)\right]
\eea
with the integrated splitting functions
\bea\label{IntSplit}
\Gamma_q(q,Q) &=& \frac{2C_F}{\pi}\frac{\alpha_s(q)}{q}
                  \left(\log\frac{Q}{q} - \frac34\right)\;,\nnb\\
\Gamma_g(q,Q) &=& \frac{2C_A}{\pi}\frac{\alpha_s(q)}{q}
                  \left(\log\frac{Q}{q}-\frac{11}{12}\right)\,.
\eea
In Eqs.(\ref{SudNLLDef}) and (\ref{IntSplit}) I have explicitly
denoted the parton type -- quark (q) or gluon (g) -- and the scales
$Q$ and $q$ are now directly related to transverse momenta.
For further details I would like to refer to \cite{Ellis:1996qj}.

Now we are in the position to give expressions for $n$ jet rates 
in $e^+e^-$ annihilations in the Durham-- or $k_\perp$ scheme
\cite{Dokshitzer:1991hj,Catani:1991hj}.
In this scheme two partons belong to different jets, if
\bea\label{yij}
y_{ij} 
= \frac{2\mbox{\rm min}\{E_i^2, E_j^2\} (1-\cos\theta_{ij})}{s}
> y_{\rm jet}\,,
\eea
where the parameter $y_{\rm jet}$ regulates the "hardness" of the jets
and $s=E_{\rm cm}^2$ is the c.m. energy squared of the $e^+e^-$
pair. Defining  the jet resolution scale 
\bea
Q_{\rm jet} = \sqrt{y_{\rm jet}} E_{\rm cm}
\eea
we have 
\bea\label{NLLrates}
{\cal R}_2^{NLL} &=& 
\left[\Delta_{q}^{\rm NLL}(E_{\rm cm},Q_{\rm jet})\right]^2\nnb\\
{\cal R}_2^{NLL} &=& 
2\left[\Delta_{q}^{\rm NLL}(E_{\rm cm},Q_{\rm jet})\right]\nnb\\
&&
\times\int\limits_{Q_{\rm jet}}^{E_{\rm cm}} dq 
\frac{\Delta_{q}^{\rm NLL}(E_{\rm cm},Q_{\rm jet})}
     {\Delta_{q}^{\rm NLL}(q,Q_{\rm jet})}
\Gamma_q(q,E_{\rm cm})
\Delta_{q}^{\rm NLL}(q,Q_{\rm jet})
\Delta_{g}^{\rm NLL}(q,Q_{\rm jet})\nnb\\
&=& 
2\left[\Delta_{q}^{\rm NLL}(E_{\rm cm},Q_{\rm jet})\right]^2
\int\limits_{Q_{\rm jet}}^{E_{\rm cm}} dq 
\Gamma_q(q,E_{\rm cm})
\Delta_{g}^{\rm NLL}(q,Q_{\rm jet})
\eea
The interpretation is quite simple: Since on the parton level jet
rates are mere probabilities, the two jet rate is just the combined 
probability that neither the quark nor the anti quark have emitted
another parton at scales above the jet resolution scale. In turn the
two jet rate is a combination of two possible "histories". In either
history, one quark does not emit a parton resolvable at $Q_{\rm jet}$,
whereas the other one first propagates down to an intermediate scale
$q$ without having radiated a parton resolvable at $Q_{\rm jet}$.
Then, at $q$ it decays into a quark and a gluon and both decay
products experience an evolution down to the jet resolution scale
without branching any further. That way, the rates for higher jet
configurations can be constructed as well, for details see
\cite{Catani:1991hj}.
However, let us note that omitting the integral over $q$ differential
jet rates at NLL can be constructed, i.e.\relax \  expressions like
$d{\cal R}_3/dq$.

\subsection{Differential jet rates and the ME weight}
This leads us directly to the construction of the weights on the ME.
Having chosen the final state momenta $p_i$ according to the ME 
with $\alpha_s$ taken at the jet resolution scale the weight is
construct along the following lines:
\begin{enumerate}
\item Construct the correct "PS history":
      \begin{itemize}
      \item Merge the two partons $i$ and $j$ with the smallest
            $y_{ij}$. Combine only "allowed" pairs, like for instance 
            \{quark, gluon\} or \{quark, anti-quark\}. The momentum
            of the new parton is the sum of the momenta of $i$ and $j$. 
      \item Repeat the step above until only a $q\bar q$ pair remains.
      \end{itemize}
\item For each parton line of type $p$ between $q_{\rm in}$ and
      $q_{\rm out}$  apply a weight 
      \bea
      \frac{\Delta_{p}^{\rm NLL}(q_{\rm in},Q_{\rm jet})}
           {\Delta_{p}^{\rm NLL}(q_{\rm out},Q_{\rm jet})}\,,
      \eea
      where $q_{\rm out}$ might be $Q_{\rm jet}$ for outgoing partons.
\item For each QCD node apply a correction factor
      \bea
      \frac{\alpha_s(q_{\rm node})}{\alpha_s(Q_{\rm jet})}\,.
      \eea
\item Accept or reject the kinematical configuration according to the 
      combined weight. 
\end{enumerate}

\piccaptioninside{\label{3jetEx}}
\parpic{
\unitlength 1mm
\begin{picture}(50,40)
\put(0,0){\includegraphics[height=4cm]{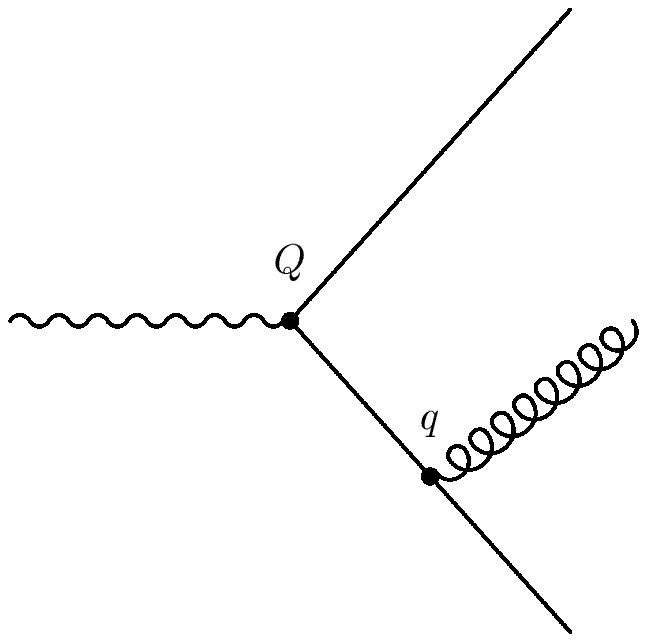}}
\end{picture}}
As an example let us consider three jet production as in
Fig.\ref{3jetEx}. In this case the correction weight reads
\bea
{\cal W} &=& \Delta_q(Q_{\rm jet},Q)
             \frac{\Delta_q(Q_{\rm jet},Q)}{\Delta_q(Q_{\rm jet},q)}
             \frac{\alpha_s(q)}{\alpha_s(Q_{\rm jet})}\nnb\\
          &&\Delta_q(Q_{\rm jet},q) 
            \Delta_g(Q_{\rm jet},q)\nnb
\eea
yielding the correct jet rate ${\cal R}_3$ after taking into account
configurations where the gluon gets clustered with the other quark
line and after inserting the approbriate integrated splitting function
and integration over all scales $q$. 

In this way, all the integrated splitting functions of the
differential jet rates at NLL accuracy have been replaced by the
correct ME. This leads to an improved description of the
region $y_{qg}\;,\;\;y_{\bar qg} > y_{\rm jet}$, i.e.\relax \  the region of
hard jet emission where interference effects matter more and more.
Clearly the combined weight of this procedure and the ME
takes into account the full ME for jet production at tree-level plus all
the higher order leading and next-to leading logaritmic contributions.

From the considerations above we can formulate the re-weighting
procedure as a cookbook recipe as follows:
\begin{enumerate}
\item Expand the PS weight up to the perturbative order in $\alpha_s$
      for jet production at tree level. In fact this boils down to 
      a product of integrated splitting functions.
\item Replace this leading term of the result with the full ME squared.    
\end{enumerate}
Let me note in passing that this is exactly the approach utilised by
{\tt Pythia} \cite{Sjostrand:2000wi} and {\tt Herwig}
\cite{Marchesini:1992ch} but only up to the first order in
$\alpha_s$. In both programmes, either the first or the hardest
emission off both the quark and the anti-quark line is re-weighted
with the ME 
\footnote{the weight is given by 
          $d\sigma_{\rm ME}(q\bar qg)/d\sigma_{\rm PS}(q\bar qg)$}, 
thus suitably replacing the splitting function. The re-weighting
procedure makes use of the fact, that 
$d\sigma_{\rm ME}(q\bar qg) < d\sigma_{\rm PS}(q\bar qg)$ in the full
phase space of gluon emission. This relation is not valid for higher
parton configurations any more and consequently the re-weighting
procedure has to be modified. The most obvious way to keep the
re-weighting applicable is to enhance the value of $\alpha_s$ in the
Sudakov form factor, $\alpha_s\to k\cdot \alpha_s$ with $k>1$ a
suitable constant factor. However, due to a potentially large number
of rejected emissions, this modification might lead to tremendously
inefficient parton showers. For more details on the rejection
procedure in various processes see
\cite{Seymour:1995df}-\cite{Corcella:1998rs}. 

\subsection{Vetoing the PS}
Having corrected the ME on differential jet rates exact up to NLL
accuracy, we are now left with the task of performing the PS. The
question then naturally arises, at which scale to start the PS.
Naively one might try to start the parton shower at the jet resolution
scale, i.e.\relax \  at $Q_{\rm jet}$. However, this is wrong. A naive counter
argument is that in such a case, there would be a radiation dip at
scales just below $Q_{\rm jet}$. For a more formal argument, consider
two jet events. Their rate is 
\bea
{\cal R}_2^{NLL}(Q_{\rm jet}) &=& 
\left[\Delta_{q}^{\rm NLL}(E_{\rm cm},Q_{\rm jet})\right]^2\,,\nnb\\
\eea
see Eq.(\ref{NLLrates}). The probability for a two jet event at an
even smaller resolution scale $Q_0<Q_{\rm jet}$ should read
\bea
{\cal R}_2^{NLL}(Q_0) &=& 
\left[\Delta_{q}^{\rm NLL}(E_{\rm cm},Q_0)\right]^2\,.
\eea
If we start the PS at the jet resolution scale, however, the
probability for two jet events at scale $Q_0$ would read
\bea
\left[\Delta_{q}^{\rm NLL}(E_{\rm cm},Q_{\rm jet})
      \Delta_{q}^{\rm NLL}(Q_{\rm jet},Q_0)\right]^2\,,
\eea
\piccaptioninside{\label{2jetVeto}}
\parpic[r]{
\unitlength 1mm
\begin{picture}(35,30)
\put(-5,-5){\includegraphics[height=4cm]{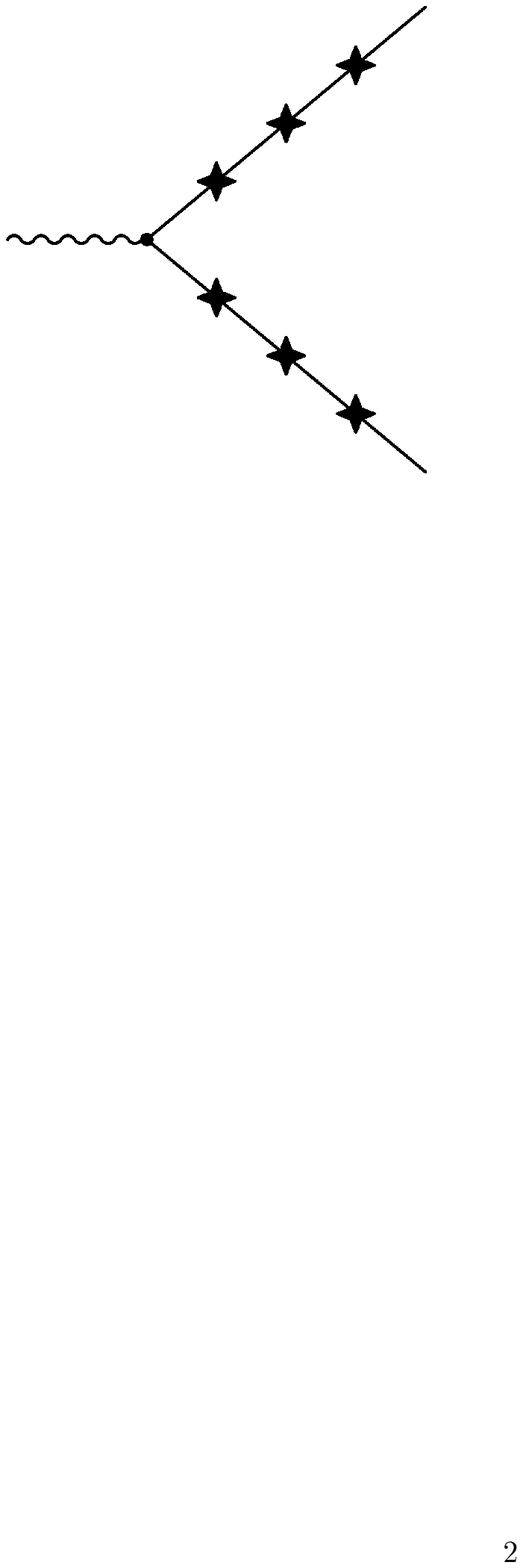}}
\end{picture}}
which is clearly wrong. The correct answer is to start the shower at
the hard scale $E_{\rm cm}$ and veto all emissions with 
$q>Q_{\rm jet}$. Identifying vetoed emissions with crosses in the
figure on the right we then find for one quark line
\bea
\Delta_q(E_{\rm cm},Q_0)\Delta_q(E_{\rm cm},Q_{\rm jet})
\left[1+\int\limits_{Q_{\rm jet}}^{E_{\rm cm}} dq
        \;\Gamma_q(q,E_{\rm cm})+\dots\right]\nnb\\
\Longrightarrow\displaystyle{
\frac{\Delta_q(E_{\rm cm},Q_0)\Delta_q(E_{\rm cm},Q_{\rm jet})}
     {\Delta_q(E_{\rm cm},Q_{\rm jet})}} = 
\Delta_q(E_{\rm cm},Q_0)\,,
\eea
exactly as demanded. Similar reasoning holds true for more jets, again
I want to refer to \cite{Catani:2001cc}. This leads to the following
algorithm for the PS:
\begin{itemize}
\item The PS evolution for {\bf any} outgoing parton starts at
      the nodal scale $q>Q_{\rm jet}$ where the parton was produced. 
\item In any branch at some nodal value $q$, the softer of the two 
      outgoing partons is produced at $q$ whereas the harder parton is
      produced at some larger scale. This is of specific interest for
      $g\to gg$ nodes. 
\end{itemize}

Taken everything together it can be shown that in the combination of 
re-weighting the MEs and vetoing the PS correct NLL jet rates are
reproduced at {\bf all} resolution scales above the fragmentation
scale. In particular, the dependence on $Q_{\rm jet}$ cancels at
NLL accuracy. Furthermore, colour configurations are chosen in a gauge
independent way. 
 
\section{ME+PS for hadron collisions}
For the construction of a combination procedure with similar virtues
we employ the same strategy of dividing the phase space for parton 
emission into a region of jet production and a region of jet evolution. 
Again, this division is achieved by means of a $k_\perp$--measure
\cite{Catani:1992zp,Catani:1993hr}, and again MEs together with
a suitable weight are responsible for the production of jets whereas
the PS together with a veto on the  emergence of extra jets takes care
of the jet evolution down to the fragmentation scale.

\subsection{Some brief reminders}
Before I go into more detail I'd like to review briefly the PS in the
initial state. For definiteness let us constrain ourselves on the case
of processes with hadrons as initial states, since the extension to
processes involving both hadrons and leptons in the initial state
(like for instance Deep--Inelastic Scattering, DIS) is
straightforward. First, some process $h_1h_2->X$ and its configuration
in phase space is generated according to 
\bea\label{tot}
\sigma_{h_1h_2\to X} =  \sum\limits_{p_1,p_2}
\int dx_1dx_2d\hat\Phi f_{p_1}^{h_1}(x_1,Q^2) f_{p_2}^{h_2}(x_2,Q^2)
                     \frac{d\hat\sigma_{p_1p_2\to X}}{d\hat\Phi}\,,
\eea
where $x_i$ are the energy fractions the partons $p_i$
carry. Furthermore, $d\hat\sigma/d\hat\Phi$ is the partial
differential cross section of the partonic subprocess w.r.t.\relax \ the
differential phase space element $\hat\Phi$ of the outgoing particles.
The scale $Q^2$ entering the parton distribution functions $f_p^h$
(PDF) depends on the phase space point under consideration. In
Drell-Yan processes, for instance, this scale is given by the
invariant mass of the lepton pair, whereas in QCD-type subprocesses
this scale is given by the $p_\perp^2$ of the outgoing
partons.

Starting from this scale the PS in the initial scale now proceeds
backwards down to some infrared scale $t_0$. This backward evolution
is due to the fact that in difference to the final state PS here the
parton configurations both at the hard and the soft scales are already
known. The idea behind the backward evolution is to include the PDFs
in order to generate only physically meaningful parton ensembles during
the PS evolution. 

In this framework, the probability for a parton $b$ to be evolved
backwards from $(t_2, x_2)$ to $(t_1,x_2)$ with no branching
resolvable at the scale $t_0$ reads
\bea
\Pi(t1,t2;x_2) = \displaystyle {
               \frac{f_b(x_2,t_1)}{f_b(x_2,t_2)}
               \frac{\Delta_b(t_2,t_0)}{\Delta_b(t_1,t_0)}}\;.
\eea
The inclusion of the PDFs actually ensures the correct physical
behaviour at low and high values of $x_2$. Having chosen a scale
$t_1$, the corresponding value $z = x_2/x_1$ of the splitting 
$a\to bc$ has to be selected according to
\bea
\displaystyle {
\frac{\alpha_s[p_\perp(z,t_1)]}{2\pi}\,
\frac{x_2/z f_a(x_2/z,t_1)}{x_2 f_b(x_2,t_1)}}
P_{a\to bc}(z)\,.
\eea
After each branching, the system is boosted into the c.m. frame
of the two outermost incoming partons.

Let us continue the discussion now with the outline of the algorithm
I'd like to propose before I start constructing the weights on the MEs.

\subsection{Proposed algorithm}
The algoritm I propose goes as follows:
\begin{enumerate}
\item Cluster initial and final state particles backwards according to
      the longitudinal invariant $k_\perp$ scheme
      \cite{Catani:1992zp,Catani:1993hr} until a $2\to 2$ process
      remains. This clustering is achieved step by step in the 
      c.m. system of the incoming partons. In this scheme the initial
      and final state partons are included in the following fashion:
      \begin{itemize}
      \item If the two particles considered are both outgoing, their
            measure $y_{ij}$ is given by
            \bea 
            y_{ij} 
            &=&   \frac{2\mbox{\rm min}\{E_i^2, E_j^2\} 
		      (1-\cos\theta_{ij})}{\hat s} \nnb\\
	    &\to& \frac{
                  \mbox{\rm min}\{p_{\perp,i}^2, p_{\perp,j}^2\} 
		       \left[(\eta_i-\eta_j)^2 +
                             (\phi_i-\phi_j)^2 \right]}{\hat s}
            \,,\nnb
            \eea
            see Eq. (\ref{yij}). Here, $\hat s$ is the invariant mass
            squared of the outgoing particles, $p_\perp$ are their
            transverse momenta w.r.t.\relax \ the beam axis and $\eta$ and
            $\phi$ are their pseudorapidities and azimuthal angles.

            If two outgoing, i.e.\relax \  final state, particles are
            clustered, the resulting particle again is a final state
            particle with $p = p_i+p_j$. 
      \item If one of the two particles, say $j$ is one of the two 
            incoming {\bf partons} then
            \bea
            y_{ij} 
            =  \frac{2E_i^2 (1-\cos\theta_{ij})}{\hat s} 
            \to \frac{p_{\perp,i}^2}{\hat s}\,.
            \eea

            If an incoming and an outgoing particle are clustered,
            the new particle is incoming, and its momentum is
            $p = p_j - p_i$. Note that in such a case a boost is in
            order to the c.m. frame of the new pair of two incoming
            particles.
      \end{itemize}
      In this scheme, the outgoing and incoming particles are
      clustered ``towards`` the hard $2\to 2$ process, i.e.\relax \ clustering
      two particles leads to another particle resolved at a higher scale.

      Note that in case we consider DIS--like processes the measures
      are given by the energies and the cosines, whereas in case
      of purely hadronic initial states the measures are given in
      terms of transverse momenta.
\item Find the hardest $k_\perp^2$ in the ``core'' $2\to 2$--subprocess.
      Examples for this hardest $k\perp^2$ are:
      \begin{itemize}
      \item $\hat s = M_{ll}^2$ in Drell-Yan type $q\bar q\to l\bar l$
            subprocesses.
      \item $\frac{2\hat s\hat t\hat u}{\hat s^2+\hat t^2+\hat u^2}$ 
            in QCD subprocesses.
      \end{itemize}
\item Apply a weight constructed by comparing histories. Some examples
      will be given below.
\item Start the initial and final state parton showers from the 
      $2\to 2$--subprocess with corresponding starting conditions,
      i.e.\relax \  every leg starts its evolution at the scale where it was
      produced.
\item Veto on jet emissions in the PS.
\end{enumerate}
 
\subsection{Constructing the ME weight}

\begin{center}
\includegraphics[width=6cm]{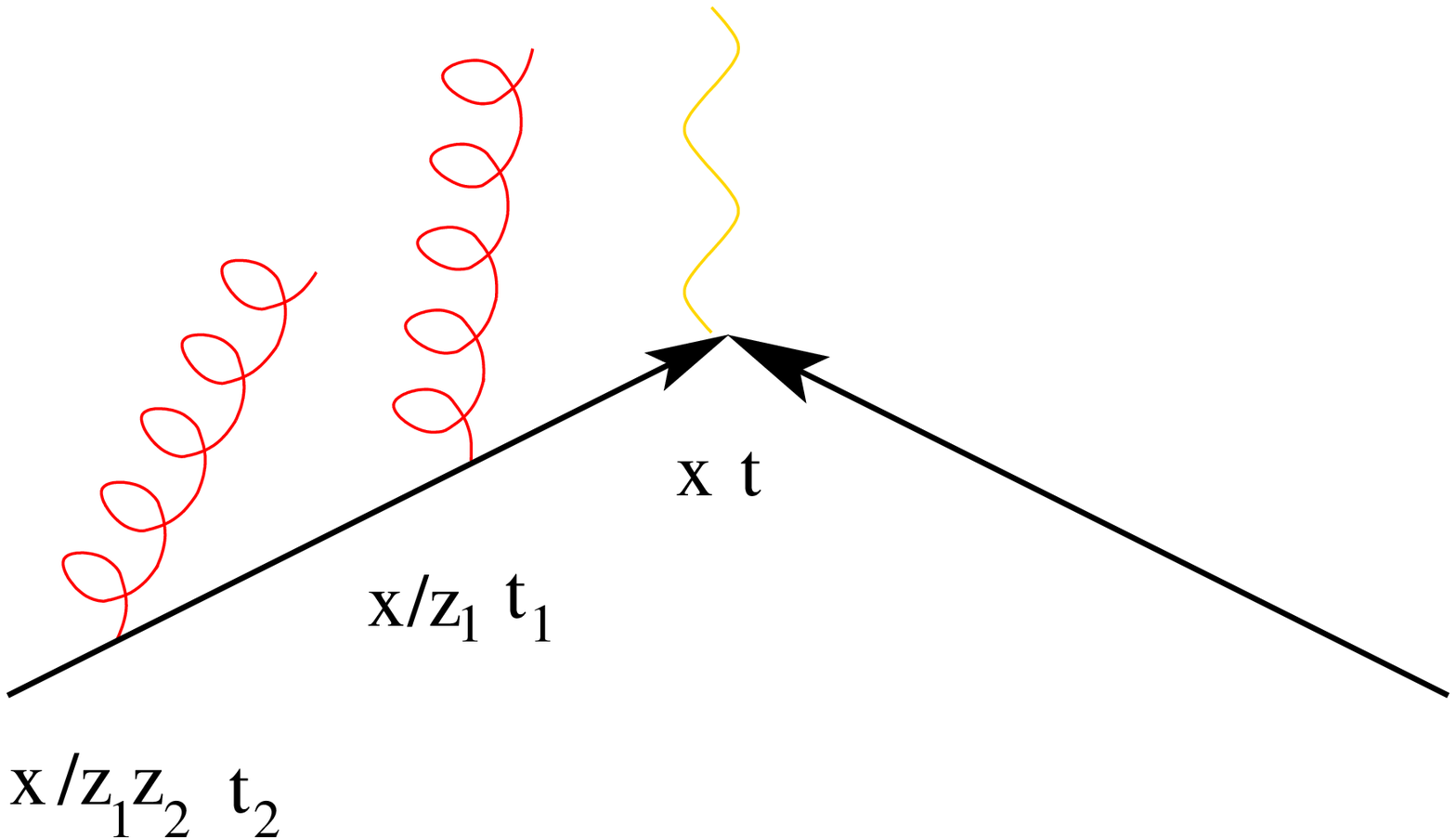}
\end{center}
As an example let us consider Drell-Yan processes. Similar to the
case of $e^+e^-$ annihilations we first construct a PS weight, that
consists of ratios of Sudakov form factors and splitting functions.
Taking $t_{\rm jet}$ as the jet resolution scale and $t$ as the hard
scale $t=M_{ll}^2$ this weight reads for the configuration in the
example above
\bea
{\cal W}_{PS} 
       &=& \;\;\;\displaystyle {
           \frac{q(x_1/z_1,t_1)}{z_1 q(x_1,t)}
           \frac{\Delta_q(t,t_{\rm jet})}{\Delta_q(t_1,t_{\rm jet})}}
           \;\cdot\alpha_s(t_1) P_{q\to qg}(z_1) \nnb\\
       &&  \times\;\;\displaystyle {
           \frac{q(x_1/z_1z_2,t_2)}{z_2 q(x_1/z_1,t_1)}
           \frac{\Delta_q(t_1,t_{\rm jet})}{\Delta_q(t_2,t_{\rm jet})}}
           \;\cdot\alpha_s(t_2)  P_{q\to qg}(z_2)\nnb\\
       &&  \times\;\;\displaystyle {
           \frac{q(x_1/z_1z_2,t_{\rm jet})}{q(x_1/z_1z_2,t_2)}
           \frac{\Delta_q(t_2,t_{\rm jet})}
                {\Delta_q(t_{\rm jet},t_{\rm jet})}}\,.
\eea
We have to combine this with the PDFs contained in the expression for
the cross section for pure Drell-Yan processes with $Q^2=t$, 
\bea
d\sigma_{DY} = 
\int dx_1dx_2 q(x_1,Q^2) \bar q(x_2,Q^2)
     d\hat\sigma_{q\bar q\to ll}(\hat s,\alpha_s(Q^2))\,,\nnb
\eea
see Eq. (\ref{tot}) and compare the result with a similar expression
for the production of two additional gluons at a lower scale,
\bea
d\sigma_{DY+gg} = 
\int dx'_1dx_2 q(x'_1=x_1/z_1z_2,t_{\rm jet}) \bar q(x_2,t_{\rm jet})
     d\hat\sigma_{q\bar q\to llgg}(\hat s/z_1z_2,\alpha_s(t_{\rm jet})\,.\nnb
\eea

We can easily read off the correction weight on the matrix element,
namely
\bea
{\cal W}_{\rm corr.} = \Delta_q(t,t_{\rm jet})
\cdot\frac{\alpha_s(t_1)}{\alpha_s(t_{\rm jet})}
\cdot\frac{\alpha_s(t_2)}{\alpha_s(t_{\rm jet})}
\eea
Note that the missing factor $z_1z_2$ in the denominator as well as
the missing PDF in the numerator is compensated by the differential
cross section for the "Drell-yan + 2 gluon"-process at the jet
resolution scale.  

\begin{center}
\includegraphics[width=6cm]{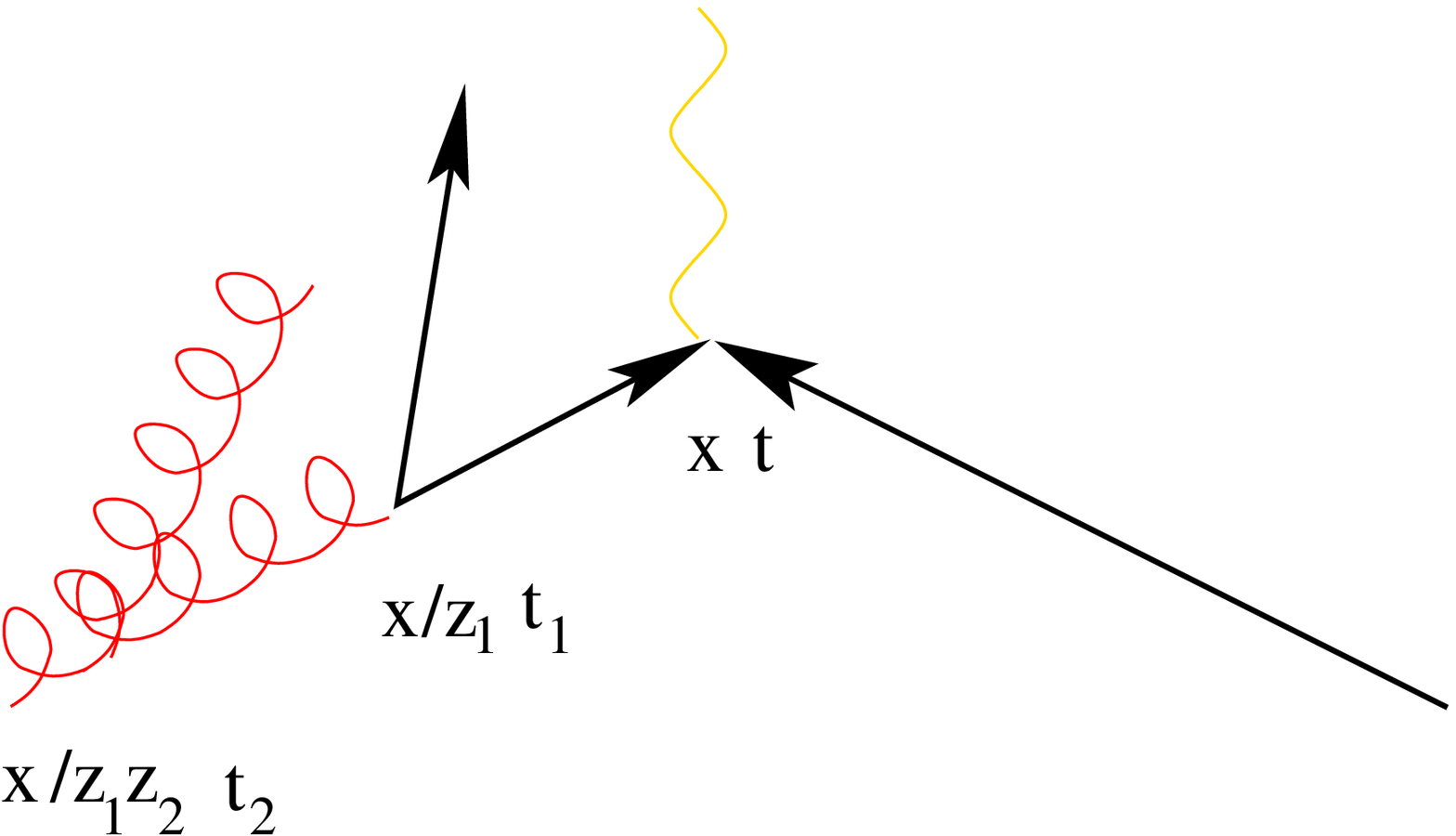}
\end{center}
Of course, things become more complicated, if some flavour changing
branchings like for instance gluon splitting occurs between the jet
resolution and the hard scale. For the example above the PS weight
would read 
\bea
{\cal W}_{PS} 
       &=& \;\;\;\displaystyle {
           \frac{g(x_1/z_1,t_1)}{z_1 q(x_1,t)}
           \frac{\Delta_q(t,t_{\rm jet})}{\Delta_q(t_1,t_{\rm jet})}}
           \;\cdot\alpha_s(t_1) P_{g\to qq}(z_1) \nnb\\
       &&  \times\;\;\displaystyle {
           \frac{g(x_1/z_1z_2,t_2)}{z_2 g(x_1/z_1,t_1)}
           \frac{\Delta_g(t_1,t_{\rm jet})}{\Delta_g(t_2,t_{\rm jet})}}
           \;\cdot\alpha_s(t_2) P_{g\to gg}(z_2)\nnb\\
       &&  \times\;\;\displaystyle {
           \frac{g(x_1/z_1z_2,t_{\rm jet})}{g(x_1/z_1z_2,t_2)}
           \frac{\Delta_g(t_2,t_{\rm jet})}
                {\Delta_g(t_{\rm jet},t_{\rm jet})}}\,,
\eea
resulting in the following correction weight
\bea
{\cal W}_{\rm corr.} =
        \displaystyle{
        \Delta_g(t_1,t_{\rm jet})
        \cdot\frac{\Delta_q(t,t_{\rm jet})}
             {\Delta_q(t_1,t_{\rm jet})}}
	\cdot\frac{\alpha_s(t_1)}{\alpha_s(t_{\rm jet})}
	\cdot\frac{\alpha_s(t_2)}{\alpha_s(t_{\rm jet})}
\,.
\eea

We read off that the correction factor is a combination of Sudakov
form factors depending only on the nodes of emissions and the flavours
of the incoming lines. This is in fact quite similar to what we
already knew from pure final state considerations.

\subsection{Vetoed showers}
Finally, let us turn to the PS. For the final state particles, the PS
evolution is done in the same fashion as in the case of $e^+e^-$
annihilations. For the two initial state lines the evolution from the
harder -- spacelike -- scales down to lower scales proceeds as
follows:
\begin{enumerate}
\item Evolve to lower scales with help of the Sudakov form factors.
\item Veto on all emissions with $t>t_{\rm jet}$. 
\item For the first allowed branching with 
      $t_{\rm all.}\le t_{\rm jet}$ the PDF part of the weight reads:
      \bea
      {\cal W}_{\rm PDF} = 
      \frac{f(\tilde x/\tilde z,t_{\rm all})}
           {\tilde z f(\tilde x,t_{\rm jet})}\,.
      \eea
      In the two examples above, $\tilde x = x_1/z_1z_2$.       
\end{enumerate}

\section{Summary and outlook}
In this paper I have suggested a method to combine MEs and PS for
hadronic interactions. This method is an extension of the already known
one for $e^+e^-$ annihilations \cite{Catani:2001cc} that has been implemented 
and successfully tested vs. experimental data in the new event
generator {\tt AMEGIC++/APACIC++} \cite{Krauss:2001iv,Krauss:1999fc}. 

In both cases the method correctly takes into account the full
perturbative order in $\alpha_s$ of the tree--level ME for the
corresponding multi jet production. Additionally, all contributions up
to NLL accuracy are reproduced in the jet rates via a re-weighting
procedure. Suitable starting conditions and a veto applied on the PS
avoid double counting. 

Nevertheless, I would like to stress that the extension of the method
I proposed still waits for the proof of correctness and for the
implementation into an event generator. This is work in progress.

\section*{Acknowledgements}
It is a pleasure for me to thank Bryan Webber, Stefano Catani, Stefano
Frixione, and Ralf Kuhn for intensive and pleasant discussions about
this subject. I'm grateful for the warm hospitality at the Cavendish
laboratory, Cambridge, where part of this work was completed.
Furthermore I would like to thank the Higgs working group and
especially Elsbieta Richter-Was of the ATLAS collaboration for the
opportunity to present my ideas.  

Financial support by DAAD is acknowledged.

\end{document}